\documentclass[aip,jap,reprint,twocolumn]{revtex4-2}

\usepackage{etoolbox,graphicx,amsmath,amssymb,color}
\usepackage[utf8]{inputenc}
\usepackage[T1]{fontenc}
\usepackage[english]{babel}

\makeatletter
\def\@email#1#2{
 \endgroup
 \patchcmd{\titleblock@produce}
 {\frontmatter@RRAPformat}
 {\frontmatter@RRAPformat{\produce@RRAP{*#1\href{mailto:#2}{#2}}}\frontmatter@RRAPformat}{}{}}
\makeatother

\begin{document}

\title{Coupling between conduction and near-field radiative heat transfer\\in tip\,-\,plane geometry}

\author{Chams Gharib Ali Barura}
\affiliation{Université Paris-Saclay, Institut d'Optique Graduate School, CNRS, Laboratoire Charles Fabry, 91127, Palaiseau, France}
\author{Philippe Ben-Abdallah}
\affiliation{Université Paris-Saclay, Institut d'Optique Graduate School, CNRS, Laboratoire Charles Fabry, 91127, Palaiseau, France}	
\author{Riccardo Messina}
\email{riccardo.messina@institutoptique.fr}
\affiliation{Université Paris-Saclay, Institut d'Optique Graduate School, CNRS, Laboratoire Charles Fabry, 91127, Palaiseau, France}

\date{\today}

\begin{abstract}
We analyze the coupling between conduction and radiative heat transfer in near-field regime between two coaxial cylinders separated by a vacuum gap. By solving the heat transport equation in the steady-state regime between metals or polar materials we highlight a flux saturation mechanism for the radiative transfer even without non-local effect. In the case of polar materials this saturation occurs in the separation distances range of $1$ to $10\,$nm which can be experimentally explored.
\end{abstract}

\maketitle

Two bodies at different temperatures exchange energy even when separated by vacuum via radiative heat transfer. The associated flux is limited in the far field (for separation distances $d\gg \hbar c/k_BT$, of some microns at ambient temperature) by Stefan\,-\,Boltzmann's law. The advent of fluctuational electrodynamics~\cite{Rytov89,Polder71} showed that this limit can be largerly overcome in the near-field regime ($d\ll\hbar c/k_BT)$, where photons tunneling between the two bodies can represent the main contribution to the energy exchange, in particular when the two bodies support electromagnetic surface resonant modes, such as surface waves~\cite{Joulain05} or a continuum of evanescent modes such as hyperbolic modes~\cite{Biehs12}. This idea has triggered the discussion of a variety of possible applications~\cite{DiMatteo01,Narayanaswamy03,Basu07,Fiorino18,DeWilde06,Jones12,Challener09,Stipe10,Abdallah13a,Abdallah15,Latella21} and has been explored experimentally in a number of experiments (see Refs.~\onlinecite{Cuevas18,Song15,Biehs21} for some recent review papers), typically confirming the theoretical predictions. Nevertheless, some of these experiments have observed deviations from the results of fluctuational electrodynamics, both in the extreme-near-field regime~\cite{Kittel} (nanometer and sub-nanometer range of distances) and at tens of nanometers~\cite{ShenNanoLetters09}, whose explanation remains to date elusive. Some theoretical interpretations (mainly invoking the participation of different carriers to the heat transfer) have been put forward in the extreme near field~\cite{MessinaarXiv,Zhang18,Alkurdi20,Volokitin19,Volokitin20,Volokitin21,Tokunaga21}, but these can only apply below a few nanometers. Another issue which has been investigated and could be relevant at slightly larger distances is the coupling between conduction within each body and the radiative flux exchanged between them. After some first works on this topic~\cite{JoulainJQSRT,Chiloyan2015natcomm}, it was shown~\cite{Messina2016} that for two parallel slabs in the diffusive regime (thickness much larger than the mean free path of phonons) the conduction\,-\,radiation coupling can induce a temperature profile within each body along with a saturation of the exchanged flux. A more general study for arbitrary transport regimes~\cite{Reina2021a} has confirmed that this effect is strong in the diffusive regime and can also strongly impact the thermalization dynamics of close bodies~\cite{Reina2021b}. The aim of the present work is to go beyond the configuration of two parallel slabs and explore the impact of geometry on the conduction\,-\,radiation coupling. More specifically, we are going to investigate a geometry allowing us to simulate the experimentally interesting configuration of a sharp tip in front of a planar substrate. 

\begin{figure}
	\centering
	\includegraphics[width=0.45\textwidth]{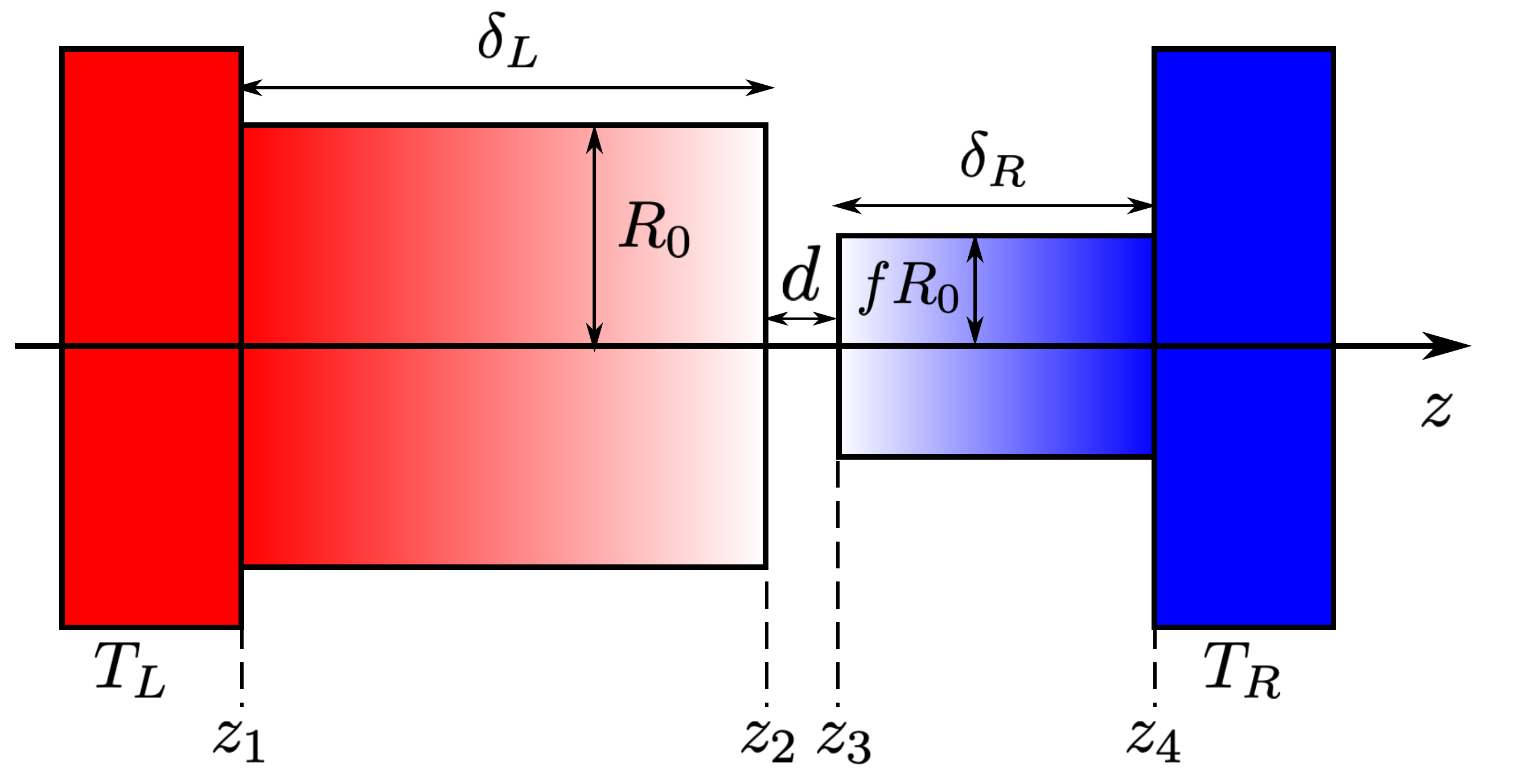}
	\caption{Sketch of the geometry of the system.}\label{fig:scheme}
\end{figure}

To start we consider two coaxial cylinders as sketched in (Fig.~\ref{fig:scheme}). The left (right) one occupies the region $z_1<z<z_2$ ($z_3<z<z_4$) and have a radius $R_0$ ($fR_0$, with $f\leq1$). The two cylinders are aligned and separated by a vacuum gap of thickness $d=z_3-z_2$. We assume that the temperatures at the external boundaries $z=z_1$ and $z=z_4$ are kept constant at $T_L$ and $T_R$, respectively, by two thermostats. Our purpose is to solve the heat equation in steady state regime $\nabla(\kappa \nabla T) = -\dot{Q}$, where $\kappa$ is the thermal conductivity of solids is assumed to be the same in the two cylinders, $T$ is the temperature profile, and $\dot{Q}$ a volumetric heat source stemming from the radiative heat transfer between the two cylinders. Taking into account the fact that in the near field the Poynting vector decays over a very thin layer of some nanometers inside each body, we will assume, as done in Ref.~\onlinecite{Messina2016}, that the radiative heat transfer can be modeled as a surface boundary condition, allowing us to reduce the heat equation to a source-free equation. The solution of this equation reads in cylindrical coordinates
\begin{equation}
	T(r,z) = a_0 + b_0 z + \sum_{k=1}^{\infty}\Bigl(a_k e^{\alpha_k\frac{z}{R}} + b_k e^{-\alpha_k\frac{z}{R}}\Bigr)J_0\Bigl(\alpha_k\frac{r}{R}\Bigr),
\end{equation}
where $J_n(x)$ is the $n$-th order Bessel function of the first kind, $\{\alpha_i\}$ ($i=0,1,\dots$) is the set of zeros of $J_1(x)$ and the coefficients $\{a_i,b_i\}$ for $i=0,1,\dots$ have to be determined. The externally imposed temperatures translate into $T(r,z_1)=T_L$ and $T(r,z_4)=T_R$, whereas in the internal boundaries $z=z_2,z_3$ we impose the continuity of the flux $\partial T(r,z_i)/\partial r=-\varphi_i(r)/\kappa$, where $i=2,3$ and $\varphi_i(r)$ represents the surface energy flux on the surface at $z_i$. At this stage, we add a further approximation, inspired by the Derjaguin approximation~\cite{DerjaguinQuartRev68}, widely used when modeling near-field radiative heat transfer in complex geometries. We assume that heat is exchanged only between points of two surfaces $z=z_2,z_3$ having the same $r$ coordinate. This implies that the surface flux $\varphi_2(r)$ at $z=z_2$ vanishes for $r>fR_0$, while for $r\leq f R_0$ we have $\varphi_2(r)=\varphi_3(r)=\varphi(r)$. For the expression of $\varphi(r)$, we employ the calculation of the flux between two planar parallel slabs.~\cite{BenAbdallah10,Biehs10} $\Phi(T(r,z_2),T(r,z_3),d)$ at the local temperatures $T(r,z_2)$ and $T(r,z_3)$ and placed at distance $d$ between each other.

We first assume that the flux takes the form $\varphi(r)=\Phi(T(r=0,z_2) , T(r=0,z_3),d)$, i.e. only depends on the temperatures at the center of the two inferfaces facing each other. This is equivalent to neglecting any $r$ dependence of the temperature in the right cylinder. This allows us to solve the coupled heat equations analytically, obtaining for the smaller cylinder ($z_3<z<z_4$) the linear profile
\begin{equation}\label{TR0}
	T(r, z) = T_R + \frac{\Phi(T(0,z_2) , T(0,z_3),d)}{\kappa} (z_4 - z),
\end{equation}
and for the larger one ($z_1<z<z_2$)
\begin{equation}\label{TL0}\begin{split}
		T(r,z) &= T_L - f^2\frac{\Phi(T(0,z_2) , T(0,z_3),d)}{\kappa} (z-z_1)\\
		&\,- 2f R_0\frac{\Phi(T(0,z_2) , T(0,z_3),d)}{\kappa}\\
		&\,\times\sum_{k=1}^{\infty}\frac{J_1(f\alpha_k)}{\alpha_k^2 J_0^2(\alpha_k)}\frac{\sinh\Bigl[\frac{\alpha_k(z - z_1)}{R_0}\Bigr]}{\cosh\Bigl(\frac{\alpha_k\delta_L}{R_0}\Bigr)}J_0\Bigl(\alpha_k\frac{r}{R}\Bigr).
\end{split}\end{equation}
Note that this solution is still implicit, since the right-hand sides of both Eqs.~\ref{TR0} and \ref{TL0} depend on $T(0,z_2)$ and $T(0,z_3)$. A fully closed-form solution can be obtained by introducing, as already done in Ref.~\onlinecite{Messina2016}, the following explicit form of the exchanged flux
\begin{equation}\label{SimplFlux}
	\Phi(T(0,z_2) , T(0,z_3),d)\simeq\frac{\gamma[T(0,z_2) - T(0,z_3)]}{d^\beta},
\end{equation}
where we linearized the flux with respect to the temperature difference and have assumed a power law scaling $d^{-\beta}$ with respect to the distance~\cite{Joulain05}. For polar materials in slab\,-\,slab configuration we have $\beta=2$. By evaluating Eqs.~\ref{TR0} and \ref{TL0} at $r=0$ and $z=z_3,z_2$, respectively, by subtracting them and coupling this to Eq.~\ref{SimplFlux} we obtain the full closed-form analytical expression of the temperature profile in the two cylinders
\begin{widetext}
\begin{equation}\label{Texact}
	T(r, z) = \begin{cases}
		T_L - \frac{\gamma(T_L - T_R)}{\xi}\Biggl[f^2(z-z_1)+2 R_0f\sum_{k=1}^{\infty}\frac{J_1(f\alpha_k)}{\alpha_k^2 J_0^2(\alpha_k)}\frac{\sinh\Bigl[\frac{\alpha_k(z - z_1)}{R_0}\Bigr]}{\cosh\Bigl[\frac{\alpha_k(z_2 - z_1)}{R_0}\Bigr]}J_0\Bigl(\alpha_k\frac{r}{R}\Bigr)\Biggr], & z_1<z<z_2,\\
		T_R + \frac{\gamma(T_L - T_R)}{\xi} (z_4 - z), & z_3<z<z_4,\end{cases}
\end{equation}
\end{widetext}
where
\begin{equation}\begin{split}
\xi&=\kappa d^2+\gamma(f^2\delta_L+\delta_R)+2\gamma R_0f\Gamma\Bigl(f,\frac{\delta_L}{R_0}\Bigr),\\
\Gamma(f,\beta)& = \sum_{k=1}^{\infty}J_1(f\alpha_k)\tanh(\alpha_k\beta)/[\alpha_k^2 J_0^2(\alpha_k)],
\end{split}
\end{equation}
along with the one of the exchanged flux
\begin{equation}\label{eq:phid}
\varphi(d,f,R_0) = \frac{\frac{\gamma\kappa (T_L - T_R)}{\kappa d^2}}{1+\frac{\gamma}{\kappa d^2}\Bigl[f^2\delta_L+\delta_R+2R_0f\Gamma\Bigl(f,\frac{\delta_L}{R_0}\Bigr)\Bigr]},
\end{equation}
$\delta_L=z_2-z_1$ ($\delta_R=z_4-z_3$) being the height of the larger (smaller) cylinder. Note that Eq.~\eqref{eq:phid} allows to recover the diverging flux $\Phi(d)=\gamma(T_L-T_R)/d^2$ in the absence of conduction\,-\,radiation coupling, corresponding to $\kappa\to\infty$, and the results for two parallel slabs~\cite{Messina2016} for $f=1$ [note that $\Gamma(1,x)=0$].

\begin{figure}
	\centering
	\includegraphics[width=0.45\textwidth]{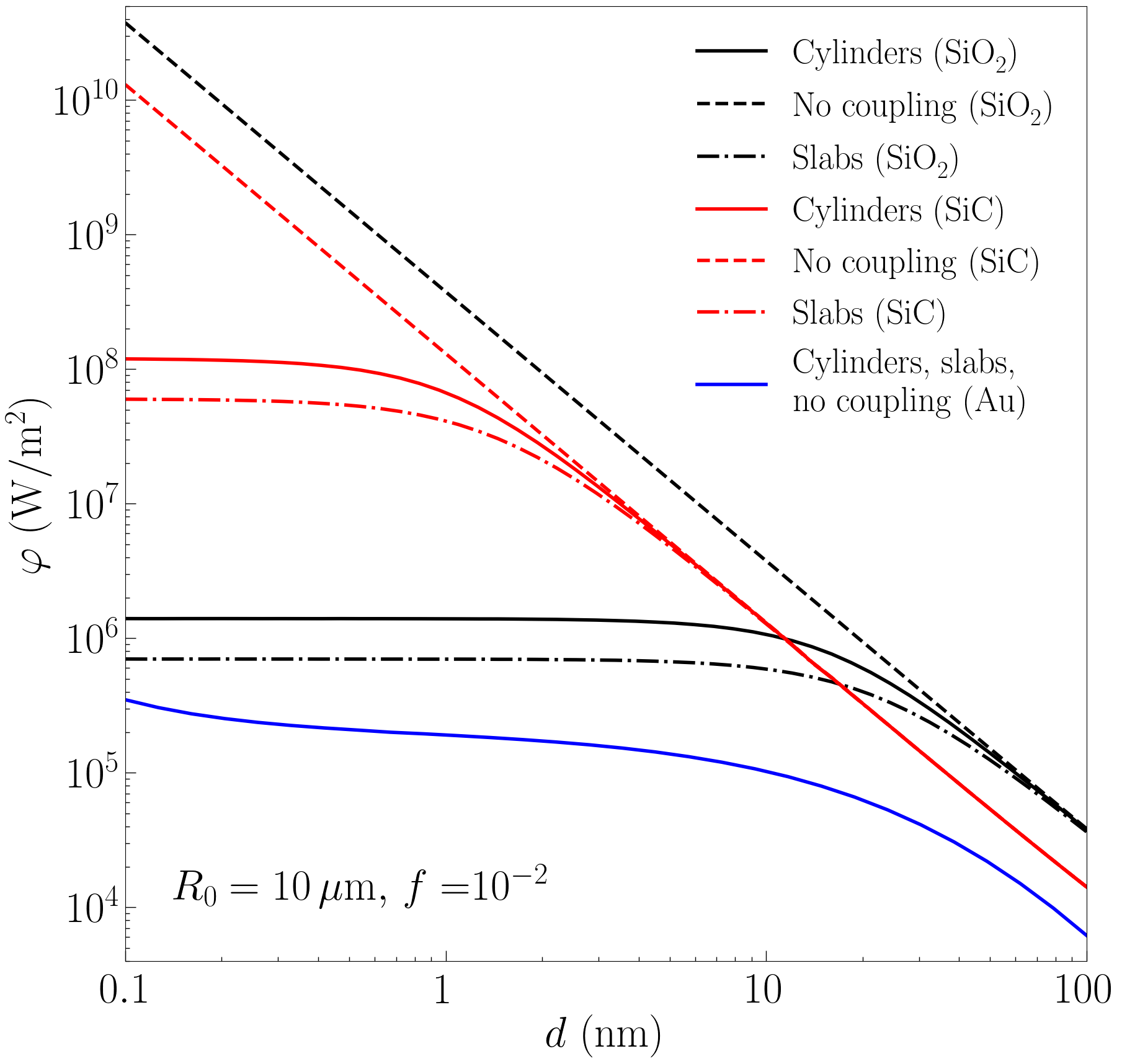}
	\caption{Radiative heat flux as a function of $d$ for $R_0=10\,\mu$m and $\delta_L=\delta_R=100\,\mu$m. The solid line corresponds to a tip\,-\,plane configuration ($f=10^{-2}$), the dot-dashed line to the slab-slab scenario ($f=1$), whereas the dashed line accounts for the $d^{-2}$ divergence (absence of coupling). The different colors correspond to different materials: black for SiO$_2$, red for SiC and blue for Au, for which the effect of coupling is not observable.}\label{fig:phid}
\end{figure}

Figure \ref{fig:phid} shows the radiative heat flux for $(T_L,T_R)=(400,300)\,$K (fixed throughout the paper) for two polar materials, silicon dioxide (SiO$_2$) and silicon carbide (SiC). Optical data are taken from Ref.~\onlinecite{Palik98}, the conductivities are $\kappa_{\text{SiO}_2}=1.4\,$W/$\text{m}\cdot\text{K}$ and $\kappa_{\text{SiC}}=120\,$W/$\text{m}\cdot\text{K}$, and the values of $\gamma$ (see Eq.~\eqref{SimplFlux}) stem from a conventional fluctuational-electrodynamics calculation, yielding $\gamma_{\text{SiO}_2}=3.8\times10^{-12}\,$W/K and $\gamma_{\text{SiC}}=1.3\times10^{-12}\,$W/K. We have taken $R_0=10\,\mu$m and $f=10^{-2}$, resulting in a tip radius $fR_0=100\,$nm. The solid black and red lines (SiO$_2$ and SiC, respectively) clearly highlight a saturation effect already observed for two parallel slabs~\cite{Messina2016} ($f=1$, dot-dashed lines in Fig.~\ref{fig:phid}). Compared to this configuration, while the deviation with respect to the theoretical $d^{-2}$ divergence (absence of coupling) takes place in the same distance range, the flux saturates to higher values in the case of two cylinders, in agreement with the lower flux exchanged through radiation. Moreover, we observe that the flux is higher for SiO$_2$ with respect to SiC in the absence of coupling (dashed lines), while the opposite is true in the presence of coupling, both in a tip\,-\,plane and a slab\,-\,slab configurations. This is a consequence of a stronger conduction\,-\,radiation coupling for SiO$_2$, stemming from a higher radiative flux ($\gamma_{\text{SiO}_2}>\gamma_{\text{SiC}}$) and a lower thermal conductivity ($\kappa_{\text{SiO}_2}<\kappa_{\text{SiC}}$). More quantitatively, while the coupling should be observable in the nanometer range for SiC, this distance range moves to tens of nanometers for SiO$_2$. An estimate of the characteristic distance at which this deviation becomes relevant is given by $\tilde{d}=\sqrt{2\gamma\delta/\kappa}$, at which for two slabs the flux is halved with respect to the absence of coupling ($\varphi(\tilde{d},f=1,R_0)=\Phi(d)/2$). This gives $\tilde{d}_{\text{SiO}_2}=23$\,nm for SiO$_2$ and $\tilde{d}_{\text{SiC}}=1.5$\,nm for SiC, in agreement with the numerical results.

We have also performed our analysis by using the full expression of the radiative heat flux at distance $d$ stemming from fluctuational electrodynamics, i.e by abandoning both the linearization with respect to $T(0,z_2)-T(0,z_3)$ and the $d^{-2}$ scaling of the flux. By evaluating Eqs.~\eqref{TR0} and \eqref{TL0} at $r=0$ and $z=z_3,z_2$, respectively, and by injecting these expressions in $\Phi(T(0,z_2) , T(0,z_3),d)$, we easily obtain the non-linear equation $\Phi(T_L-ax,T_R+bx) - x = 0$, where $a$ and $b$ can be easily obtained from Eqs.~\eqref{TR0} and \eqref{TL0}, while $x$ represents the flux $\varphi(T(0,z_2) , T(0,z_3),d)$, from which the temperature profile in both cylinders can be deduced. The solution of this equation can be easily found numerically, yielding for any separation distance $d$ the full temperature profiles and exchanged flux in the presence of coupling. This approach allowed us to verify the validity of Eq.~\eqref{SimplFlux}, obtaining curves which are indistinguishable from those shown in Fig.~\ref{fig:phid} for polar materials. Moreover, it allowed us to investigate the role of coupling in the case of metals, for which the $d^{-2}$ scaling law for the distance dependence of flux is not valid anymore. More specifically, we have considered two gold (Au) cylinders whose dielectric constant is described by means of a Drude model~\cite{Palik98} and having a thermal conductivity $\kappa_{\text{Au}}=310\,$W/$\text{m}\cdot\text{K}$. Figure \ref{fig:phid} shows that the curves for gold in the absence of coupling, and in the presence of coupling both for two slabs and in the tip\,-\,plane configuration are superposed. This is, not surprisingly, a result of the high thermal conductivity of gold along with the much lower radiative flux due to the absence of surface resonant modes in the infrared region of the spectrum.

Finally, we have verified the validity of the approximation neglecting the $r$ dependence of the flux, by replacing Eq.~\eqref{SimplFlux} by the expression $\varphi(r)\simeq\gamma[T(r,z_2) - T(r,z_3)]/d^2$. By solving the time-independent heat equation in the presence of this $r$-dependent flux and injecting the solution of the two temperatures inside its expression we obtain the integral equation for the exchanged flux
\begin{widetext}
\begin{equation}\begin{split}
		&\varphi(r) = \frac{\gamma(T_L - T_R)}{d^2} - \frac{2\gamma}{\kappa d^2}\Biggl\{\frac{f^2\delta_L+\delta_R}{f^2R_0^2}\int_0^{fR_0}\!dr'r'\,\varphi(r')+\sum_{k=1}^{\infty}\frac{1}{\alpha_k fR_0J_0^2(\alpha_k)}\\
		&\times\int_0^{fR_0}\!dr'r'\,\Bigl[f\tanh\Bigl[\frac{\alpha_k\delta_L}{R_0}\Bigr]J_0\Bigl(\frac{\alpha_kr}{R_0}\Bigr)J_0\Bigl(\alpha_k\frac{r'}{R_0}\Bigr)+\tanh\Bigl[\frac{\alpha_k\delta_R}{fR_0}\Bigr]J_0\Bigl(\alpha_k\frac{r}{fR_0}\Bigr)J_0\Bigl(\frac{\alpha_kr'}{fR_0}\Bigr)\Bigr]\varphi(r')\Biggr\}.
\end{split}\end{equation}
\end{widetext}
For any experimentally reasonable choice of the parameters $d$, $R_0$ and $f$, the numerical solutions of this equation both for SiC and SiO$_2$ are indistinguishable from the ones of Fig.~\ref{fig:phid}, confirming the negligible dependence of the temperature profiles on the radial coordinate.

\begin{figure}
	\centering
	\includegraphics[width=0.42\textwidth]{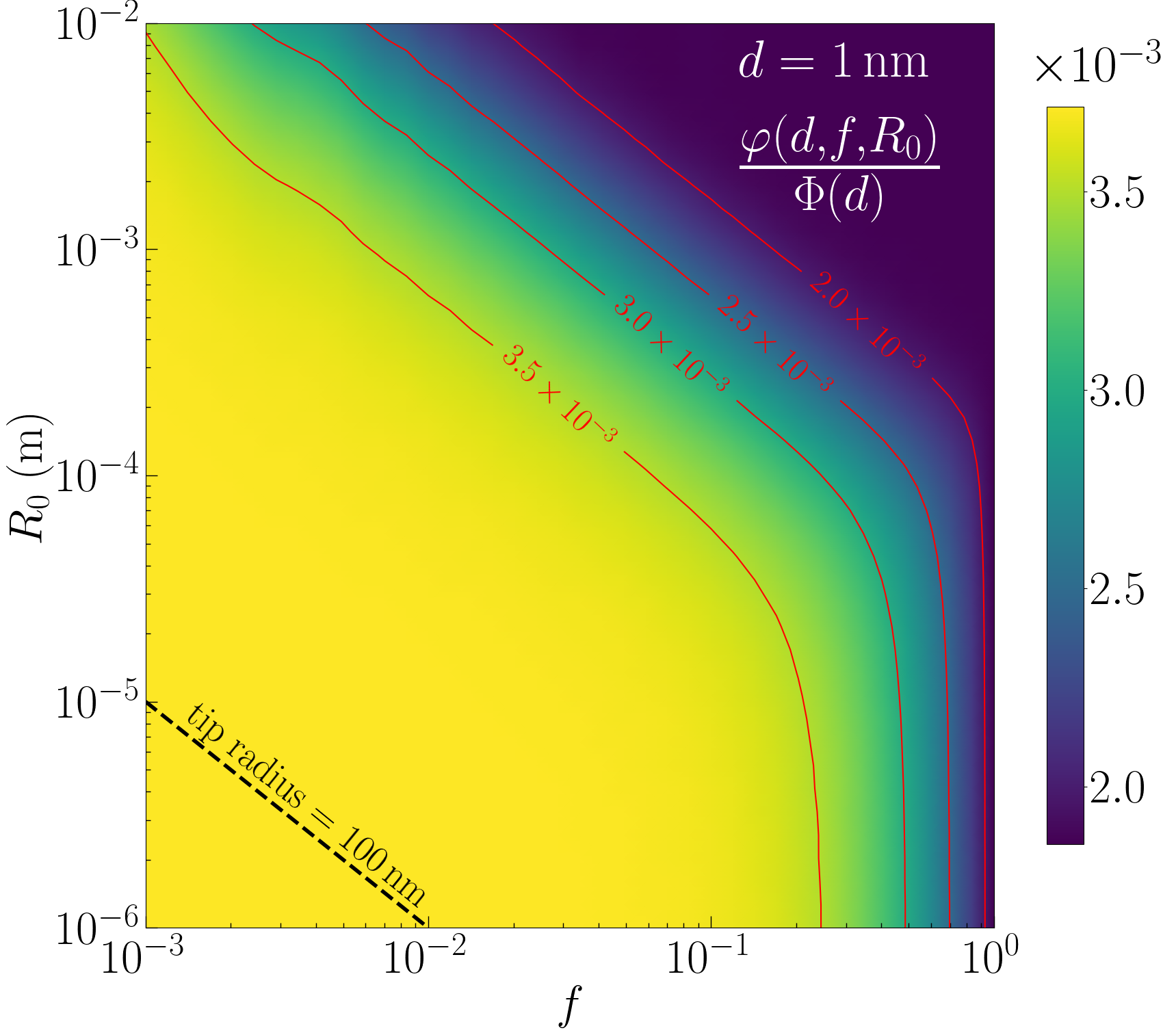}
	\caption{Ratio between the exchanged fluxes in the presence and absence of coupling as a function of the radius $R_0$ of the larger cylinder and the ratio $f$ between the two cylinders radii. The other parameters are $d=1\,$nm and $\delta_L=\delta_R=100\,\mu$m. Some contour lines are indicated in red, while the black dashed line corresponds to a fixed tip radius $fR_0=100\,$nm.}\label{fig:eta}
\end{figure}

We have used the formalism developed so far to investigate the role of geometry in the conduction\,-\,radiation coupling. For two SiO$_2$ cylinders at a distance $d=1\,$nm, Fig.~\ref{fig:eta} shows the ratio between the exchanged flux in the presence of coupling $\varphi(d,f,R_0)$ and the one in the absence of coupling $\Phi(d)$, as a function of the radius $R_0$ of the larger cylinder (representing the plane) and the ratio $f$ between the two radii.
It also highlight a reduction of the flux between two and three orders of magnitude. Moreover, albeit the large range of explored parameters, this ratio is not significantly varying, proving that we are already in a saturation regime. Furthermore, it is also interesting to look at a cut of this density plot for a given value of the tip radius $fR_0$. A value of $fR_0=100\,$nm is shown in Fig.~\ref{fig:eta}, allowing us to conclude that not only is this saturation clearly taking place for small tips, but that for those values we are safely in a saturation regime.

\begin{figure}
	\centering
	\includegraphics[width=0.4\textwidth]{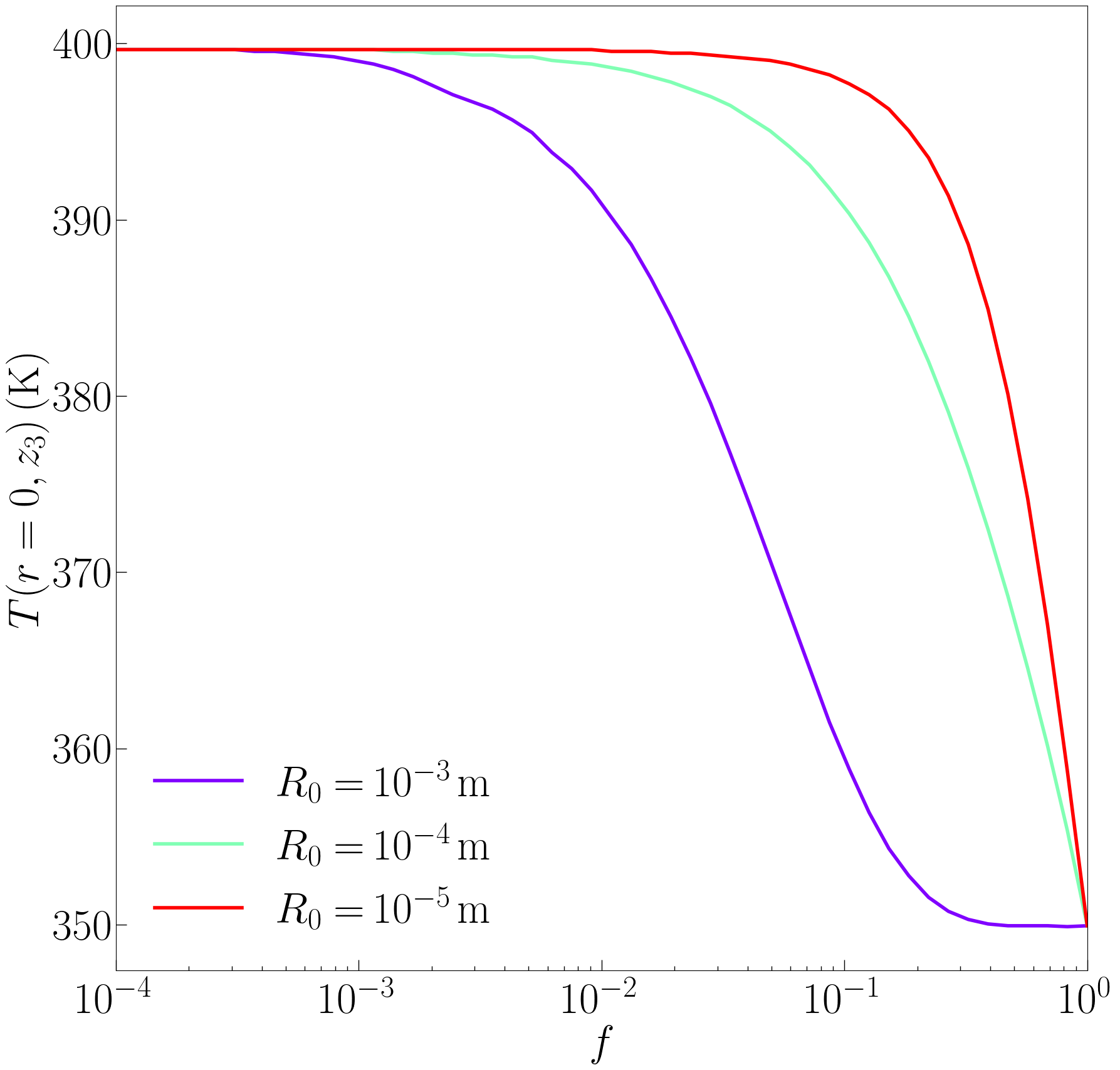}
	\caption{Temperature of the apex of the tip as a function of the radii fraction $f$, for three different values of the radius $R_0$ of the larger cylinder. The other parameters are $d=1\,$nm and $\delta_L=\delta_R=100\,\mu$m.}\label{fig:T2}
\end{figure}

We conclude by studying the temperature $T(r=0,z=z_3)$ of the apex of tip. This quantity is shown in Fig.~\ref{fig:T2} as a function of $f$ for different radii $R_0$ of the larger cylinder. We identify two limiting cases: for $f=1$ we have two parallel slabs both thermalizing, in their closest points $z=z_2,z_3$, to $350\,$K, i.e. the average of the two externally imposed temperatures $T_L$ and $T_R$. On the contrary, for $f\to0$ the apex of the tip tends to thermalize to the temperature of the left cylinder (the plane), almost independent of $z$ and equal to $T_L$. These limiting temperatures strongly depend on the material under scrutiny (and on the distance). They can be expressed analytically, within the approximations leading to Eq.~\eqref{Texact}, as $T_{f=1}=T_R+\gamma(T_L-T_R)\delta_R/[\gamma(\delta_L+\delta_R)+\kappa^2d]$ and $T_{f\to0}=T_R+\gamma(T_L-T_R)\delta_R/(\gamma\delta_R+\kappa^2d)$ for $f=1$ and $f\to0$, respectively. For SiC this gives $T_{f\to0}=364$\,K and $T_{f=1}=335$\,K , showing a present but weaker conduction\,-\,radiation coupling with respect to SiO$_2$, while in the case of gold we obtain $T(r=0,z_3)\simeq T_{f\to0}\simeq T_{f=1}\simeq T_R=300\,$K confirming the negligible conduction\,-\,radiation coupling. 

Going back to SiO$_2$ (see Fig.~\ref{fig:T2}), we observe that the transition between the limiting cases $f\to0$ and $f=1$ takes place on a characteristic scale of $f$ which clearly depends on $R_0$. Nevertheless, while the three curves depict the possible transitions from the extremely thin tip to the slab\,-\,slab configuration for different values of $R_0$, it must be pointed out that for any reasonable value of the tip radius $fR_0$ we are always very close to the limit $f\to0$, i.e. fully in a saturated regime where the tip apex thermalizes to the highest possible temperature.

In conclusion, we have studied the coupling between conductive and near-field radiative heat transfer in a system made of two different coaxial cylinders which mimic, in the limit of small ratio between the two cylinder radii, the geometric configuration of a tip\,-\,plane thermal-microscopy experiment. We have solved the coupled heat equation in this geometry, obtaining some closed-form expressions for both the temperature profiles and the exchanged flux in some approximations, whose validity has been verified numerically. We have shown that the conduction-radiation coupling can induce a saturation of heat flux exchanged between the two solids even without non-local optical response of materials. This saturation can induce a significant temperature change in the smaller cylinder (the tip), up to tens of degrees in the case of polar materials, along with a strong reduction of the flux (of more than two orders of magnitude) with respect to the scenario of absence of coupling. While this coupling has a negligible effect in the case of metals, it is relevant for polar materials and manifests itself at separation distances up to tens of nanometers, making its experimental observability feasible. Since this distance is comparable to distances where the non-local optical response of materials is likely to play a role in the transfer, these effects should also be analyzed in future works.

\section*{Author declarations}

\subsection*{Conflict of interest}

The authors have no conflicts to disclose.

\section*{Data availability}

The data that support the findings of this study are available from the corresponding authors upon reasonable request.


\begin{thebibliography}{999}
% Fluctuational electrodynamics
\bibitem{Rytov89}S.~M. Rytov, Y.~A. Kravtsov, and V.~I. Tatarskii, \emph{Principles of Statistical Radiophysics} (Springer, New York, 1989).
\bibitem{Polder71} D. Polder and M. van Hove, Phys. Rev. B \textbf{4}, 3303 (1971).
% Surface resonances
\bibitem{Joulain05} K. Joulain, J.-P. Mulet, F. Marquier, et al., Surf. Sci. Rep. \textbf{57}, 59 (2005).
\bibitem{Biehs12} S.-A. Biehs, M. Tschikin, and P. Ben-Abdallah, Phys. Rev. Lett. \textbf{109}, 104301 (2012).

% Energy harvesting
\bibitem{DiMatteo01} R.~S. DiMatteo, P. Greiff, S.~L. Finberg, et al., Appl. Phys. Lett. \textbf{79}, 1894 (2001).
\bibitem{Narayanaswamy03} A. Narayanaswamy and G. Chen, Appl. Phys. Lett. \textbf{82}, 3544 (2003).
\bibitem{Basu07} S. Basu, Y.-B. Chen, and Z.~M. Zhang, Int. J. Energy Res. \textbf{31}, 689 (2007)
\bibitem{Fiorino18} A. Fiorino, L. Zhu, D. Thompson, et al., Nat. Nanotech. \textbf{13}, 806 (2018).
% Data recording
\bibitem{Challener09} W.~A. Challener, C. Peng, A.~V. Itagi, et al., Nat. Photon. \textbf{3}, 220 (2009).
\bibitem{Stipe10} B.~C. Stipe, T.~C. Strand, C.~C. Poon, et al., Nat. Photon. \textbf{4}, 484 (2010).
% Spectroscopy
\bibitem{DeWilde06} Y. De Wilde, F. Formanek, R. Carminati, et al., Nature \textbf{444}, 740 (2006).
\bibitem{Jones12} A.~C. Jones and M.~B. Raschke, Nano Lett. \textbf{12}, 1475 (2012).
% Thermotronics
\bibitem{Abdallah13a} P. Ben-Abdallah and S.-A. Biehs, Phys. Rev. Lett. \textbf{112}, 044301 (2013).
\bibitem{Abdallah15} P. Ben-Abdallah and S.-A. Biehs, AIP Adv. \textbf{5}, 053502 (2015).
%Applications:review
\bibitem{Latella21} I. Latella, S.-A. Biehs and P. Ben-Abdallah, Optics Express 29(16) 24816-24833 (2021).
% Review papers
\bibitem{Cuevas18} J.~C. Cuevas and F.~J. García-Vidal, ACS Photonics \textbf{5}, 3896 (2018).
\bibitem{Song15} B. Song, A. Fiorino, E. Meyhofer, et al., AIP Adv. \textbf{5}, 053503 (2015).
\bibitem{Biehs21} S.-A. Biehs, R. Messina, P.~S. Venkataram, et al., Rev. Mod. Phys. \textbf{93}, 025009 (2021).
% Extreme-near-field experiments
\bibitem{Kittel} K. Kloppstech et al.,\emph{Giant heat transfer in the crossover regime between conduction and radiation}, Nat. Commun. \textbf{8}, 14475 (2017).
% Deviations
\bibitem{ShenNanoLetters09}S. Shen, A. Narayanaswamy, and G. Chen, Nano Letters \textbf{9}, 2909 (2009).
% Extreme near field
\bibitem{MessinaarXiv}R. Messina, S.-A. Biehs, T. Ziehm, A. Kittel, and P. Ben-Abdallah, arXiv:1810.02628v1.
\bibitem{Zhang18}Z. Q. Zhang, J. T. L\"{u}, and J. S. Wang, Phys. Rev. B \textbf{97}, 195450(2018).
\bibitem{Alkurdi20}A. Alkurdi, C. Adessi, F. Tabatabaei, S. Li, K. Termentzidis,and S. Merabia, Int. J. Heat Mass Transf. \textbf{158}, 119963 (2020).
\bibitem{Volokitin19}A. I. Volokitin, JETP Lett. \textbf{109}, 749 (2019).
\bibitem{Volokitin20}A. I. Volokitin, J. Phys.: Condens. Matter \textbf{32}, 215001 (2020).
\bibitem{Volokitin21}A. I. Volokitin, Phys. Rev. B \textbf{103}, L041403 (2021).
\bibitem{Tokunaga21}T. Tokunaga, A. Jarzembski, T. Shiga, K. Parg, and M. Francoeur, Phys. Rev. B \textbf{104}, 125404 (2021).
% Conduction-radiation coupling
\bibitem{JoulainJQSRT}K. Joulain, J. Quant. Spectrosc. Radiat. Transfer \textbf{109}, 294 (2008).
\bibitem{Chiloyan2015natcomm}V. Chiloyan, J. Garg, K. Esfarjani, and G. Chen, Nature Communications \textbf{6}, 6775 (2015).
\bibitem{Messina2016} R. Messina, W. Jin, and A.~W. Rodriguez, Phys. Rev. B \textbf{94}, 121410(R) (2016).
\bibitem{Reina2021a} M. Reina, R. Messina, and P. Ben-Abdallah, Phys. Rev. Lett. \textbf{125}, 224302 (2021).
\bibitem{Reina2021b} M. Reina, R. Messina, and P. Ben-Abdallah, Phys. Rev. B \textbf{104}, L100305 (2021).
% Heat flux between slabs
\bibitem{BenAbdallah10}P. Ben-Abdallah and K. Joulain, Phys. Rev. B \textbf{82}, 121419(R) (2010).
\bibitem{Biehs10}S.-A. Biehs, E. Rousseau, and J.-J. Greffet, Phys. Rev. Lett. \textbf{105}, 234301 (2010).
% Palik
\bibitem{Palik98} \emph{Handbook of Optical Constants of Solids}, edited by E. Palik (Academic Press, New York, 1998).
% PFA
\bibitem{DerjaguinQuartRev68}B. V. Derjaguin, I. I. Abrikosova, and E. M. Lifshitz, Quart. Rev. \textbf{10}, 295 (1968).

\end{thebibliography}
\end{document}